\documentclass[preprint,12pt]{elsarticle}

\usepackage{graphicx}

\usepackage{amssymb, amsmath, color, a4wide}

\biboptions{sort,compress,merge}
\journal{Computer Physics Communications}

\begin{document}
\parindent 0pt
\begin{frontmatter}

\title{Solving the Ghost-Gluon System of Yang-Mills Theory on GPUs}

\author[label1]{Markus Hopfer\corref{cor1}}
\cortext[cor1]{corresponding author}
\ead{markus.hopfer@uni-graz.at}
\address[label1]{Institut f{\"u}r Physik, Karl-Franzens Universit{\"a}t, 
Universit{\"a}tsplatz 5, 8010 Graz, Austria}

\author[label1]{Reinhard Alkofer}
\ead{reinhard.alkofer@uni-graz.at}

\author[label2]{Gundolf Haase}
\ead{gundolf.haase@uni-graz.at}
\address[label2]{Institut f{\"u}r Mathematik und wissenschaftliches Rechnen, 
Karl-Franzens Universit{\"a}t, Heinrichstra{\ss}e 36, 8010 Graz, Austria}

\begin{abstract}
We solve the ghost-gluon system of Yang-Mills theory using Graphics Processing Units 
(GPUs). Working in Landau gauge, we use the Dyson-Schwinger formalism for the 
mathematical description as this approach is well-suited to directly benefit from the 
computing power of the GPUs. With the help of a Chebyshev expansion for the dressing 
functions and a subsequent appliance of a Newton-Raphson method, the non-linear system 
of coupled integral equations is linearized. The resulting Newton matrix is generated in 
parallel using OpenMPI and CUDA\texttrademark. Our results show, that it is possible to 
cut down the run time by two orders of magnitude as compared to a sequential version of 
the code. This makes the proposed techniques well-suited for Dyson-Schwinger 
calculations on more complicated systems where the Yang-Mills sector of QCD serves as a 
starting point. In addition, the computation of Schwinger functions using GPU devices is 
studied.
\end{abstract}


\end{frontmatter}

\section{Introduction}
It is well-established by now that quantum chromodynamics (QCD) provides the necessary 
framework to describe the strong interaction among quarks and gluons. It is furthermore 
believed that confinement, which denotes the absence of color charged objects from the 
physical spectrum, origins from the gauge sector of the theory. Here, the infrared 
properties of the one-particle irreducible Green's functions are of particular interest. 
Due to the large value of the coupling in this low-energy regime, a non-perturbative 
treatment is mandatory. The Dyson-Schwinger (or generally a Green's functions) approach 
provides a continuum formulation of QCD 
\cite{Dyson:1949ha,Schwinger:1951ex,Alkofer:2000wg,Maris:2003vk,Fischer:2006ub} 
capable to describe the system over the entire momentum range. Dyson-Schwinger equations 
(DSEs) constitute a highly coupled system of non-linear integral equations, the 
equations of motion for the underlying quantum field theory. If there were the 
possibility to solve these equations self-consistently, the whole dynamics of the 
quantum system would be uncovered \cite{Haag:1992hx}. Unfortunately, each DSE 
comprehends higher order Green's functions, such that the whole system builds up to an 
infinite tower of $n$-point functions and an appropriate truncation is mandatory. 
The truncation procedure is a highly non-trivial task and one has to account for 
errors induced by the particular truncation scheme. On the other hand, and based on 
the work of \cite{vonSmekal:1997is,Zwanziger:2001kw,Lerche:2002ep}, there has also been 
substantial progress in solving the whole tower in the far infrared 
\cite{Alkofer:2004it,Fischer:2009tn,Fister:2010yw}.
\newline\newline
Initiated by Mandelstam \cite{Mandelstam:1979xd} the gluon propagator and later the 
ghost-gluon system was at the focus of many contemporary DSE studies.
Recent investigations agree on an enhancement of the ghost whereas the gluon propagator is 
suppressed in the infrared regime 
\cite{Atkinson:1997tu,Watson:2001yv,Fischer:2003zc,Aguilar:2008xm,Boucaud:2008ky,Fischer:2008uz}. 
This picture is also enforced by results obtained from lattice simulations, see, {\it
e.g.}, 
\cite{Cucchieri:2011ga,Maas:2011se} and references therein, 
or Functional Renormalization Group methods \cite{Pawlowski:2003hq}. 
Whether or not the gluon propagator exactly vanishes in the infrared 
has been discussed for quite some time
\footnote{This would correspond to an infrared singular ghost propagator, 
{\it i.e.}, a scenario in accordance with the Kugo-Ojima 
\cite{Kugo:1979gm,Nakanishi:1990qm} 
and Gribov-Zwanziger \cite{Gribov:1977wm,Zwanziger:1991gz} 
confinement criterion.}. Although this so-called scaling solution is predicted by  
continuum approaches it is, in four space-time dimensions, not seen in recent 
lattice calculations of the gluon propagator \cite{Maas:2011se}, see however 
\cite{Sternbeck:2008mv,Maas:2009se,Maas:2009ph}. 
\newline\newline
Within the last years, Graphics Processing Units (GPUs) became an essential branch in 
high performance computing. Their efficiency, i.e. the ratio between computational power 
and power consumption, makes them a reasonable alternative to conventional clusters. 
With the availability of low-cost consumer GPU devices this technology also finds its 
way into desktop computers offering the possibility to perform general purpose 
scientific and engineering computations in an until then not feasible way. However, 
the mapping of existing algorithms and/or software to the massively parallel SIMD 
architecture of the GPU is often difficult such that a restructuring of the 
algorithms/code is inevitable in order to meet the requirements of the hardware. 
In case of the ghost-gluon system only minimal modifications are needed such that the 
portation of the sequential code is a straightforward task. The main objective of this 
paper is to employ the benefits of modern GPU devices into DSE calculations. Here, 
Yang-Mills theory is not only an interesting topic on its own but serves also as a 
starting point for investigations involving fermions since the treatment of larger 
systems by incorporating additional DSEs is possible with the proposed methods. 
Compared to the sequential code, performance gains by two orders of magnitude can be 
obtained. This paper is organized as follows. In Section \ref{sec:mathintro} we 
introduce the mathematical formulation of the problem. In Section \ref{sec:numerics} 
a description of the employed numerical methods will be given, where the parallelization 
will be detailed in Section \ref{sec:GPUsection}. Finally in Section 
\ref{sec:performance} we compare the performance of the sequential code with the 
parallelized versions using CUDA\texttrademark as well as OpenMPI and, in addition, 
the computation of Schwinger functions on GPU devices will be outlined. Our summary and 
conclusions will be given in Section \ref{sec:conclusions}.
\section{The Dyson-Schwinger Equations for the Ghost-Gluon System}
\label{sec:mathintro}
\begin{figure}[ht!]
 \centering
 \includegraphics[scale=0.5]{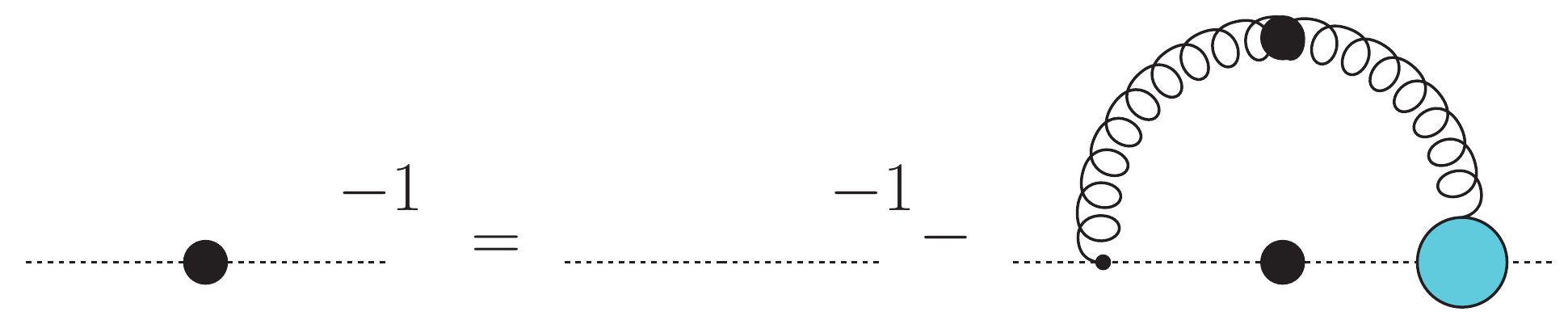}
 \caption{The DSE for the Landau gauge ghost propagator.}
 \label{fig:GhostDSE}
\end{figure}

\begin{figure}[ht!]
 \centering
 \includegraphics[scale=0.5]{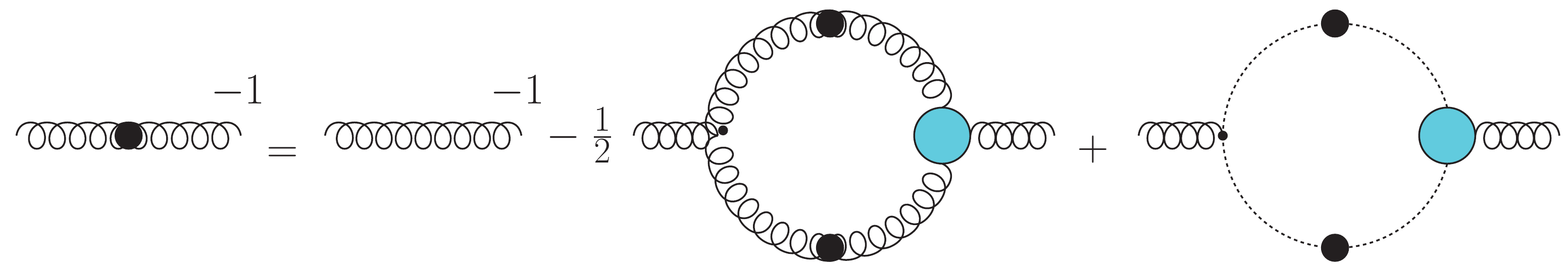}
 \caption{The truncated DSE for the Landau gauge gluon propagator.}
 \label{fig:GluonDSE}
\end{figure}

Within the DSE approach to QCD the Yang-Mills system is described by a set of coupled 
integral equations for the corresponding ghost and gluon propagator as depicted in 
Figs. \ref{fig:GhostDSE}-\ref{fig:GluonDSE}, where throughout this paper Landau gauge is used. 
Note that the DSE for the gluon propagator 
is already truncated in order to render the system tractable\footnote{Throughout this 
section the notation is adapted from \cite{Fischer:2003zc} where this system has been 
solved.}. Thus, no two-loop as well as no tadpole diagrams are taken into account. The 
black dots indicate dressed propagators whereas the blue blobs represent dressed 
vertices.

The corresponding formal expression for the ghost propagator reads
\begin{equation}
 D_G^{-1}(p) = \tilde Z_3{D_{G,0}}^{-1}(p) - 
 \tilde Z_1g^2N_c\int \dfrac{d^4q}{(2\pi)^4}\; 
 \Gamma_{\mu,0}(p,q)D_{\mu\nu}(p-q)\Gamma_\nu(q,p)D_G(q),
\end{equation}
where for the gluon propagator it is given by
\begin{equation}
 \begin{split}
  D_{\mu\nu}^{-1}(p) = Z_3D_{\mu\nu,0}^{-1}(p)\;+\;& 
  \tilde Z_1g^2N_c\int \dfrac{d^4q}{(2\pi)^4}\; 
  \Gamma_{\mu,0}(p,q)D_G(p-q)\Gamma_\nu(q,p)D_G(q) \\
   -\;& Z_1g^2\frac{N_c}{2}\int \dfrac{d^4q}{(2\pi)^4}\; 
   \Gamma_{\mu\rho\sigma,0}(p,q)D_{\rho\rho'}(p-q)\Gamma_{\rho'\nu\sigma'}(q,p)
   D_{\sigma\sigma'}(q).
 \end{split}
\end{equation}
Here, $D/\Gamma$ denotes the dressed and bare propagators/vertices respectively. 
$N_c$ is the number of colors and $\tilde Z_1$ is the renormalization constant for the 
ghost-gluon vertex which can be set to one in Landau gauge as this vertex is 
UV finite \cite{Taylor:1971ff}. Furthermore, $Z_1$ is the renormalization constant for 
the three-gluon vertex and $\tilde Z_3$ and $Z_3$ denote the wave-function 
renormalization constants for the ghost and gluon, respectively. For the purpose of this 
numerical study it is sufficient to replace the dressed ghost-gluon vertex by its bare 
counterpart, i.e. $\Gamma_\nu(q,p)=iq_\nu$, whereas for the three-gluon vertex the model 
proposed in \cite{Fischer:2003zc} is employed
\begin{equation}
 \Gamma_{\mu\nu\sigma}(q,p) = \dfrac{1}{Z_1}\dfrac{G(q^2)^{1-a/\delta-2a}}{Z(q^2)^{1+a}}
 \dfrac{G(k^2)^{1-b/\delta-2b}}{Z(k^2)^{1+b}}\Gamma_{\mu\nu\sigma,0}(q,p),
\end{equation}
with $k^2\equiv(q-p)^2$. The choice of $a=b=3\delta$, where $\delta=-9/44$ is the 
anomalous dimension of the ghost, leads to the correct scaling of the dressing 
functions in the ultraviolet region\footnote{For small momenta, the model approaches a 
constant value, which is reasonable as the gluon loop diagram is sub-leading the 
infrared regime.}. Using $O(4)$ invariance  in the Euclidean 
formulation of QCD, the four dimensional integrals can be rewritten in the form
 \begin{equation}
  \int d^4q = 4\pi\int_0^{\Lambda^2}dy\;\frac{y}{2}\int_0^\pi d\theta\sin^2\theta,
 \end{equation}
where two integrations are carried out trivially yielding a factor of $4\pi$. Here we 
introduced the abbreviation $q^2 = y$ for the loop momentum\footnote{We regularized the 
system using a sharp momentum cutoff $\Lambda^2$.}. In the following we furthermore use 
$p^2 = x$ and $k^2 = z$. For the radial integral we employ a standard 
\textit{Gauss-Legendre quadrature} rule, where the nodes are appropriately distributed 
over the whole momentum range by using a non-linear mapping function. For the angular 
integration we use a \textit{tanh-sinh quadrature} \cite{Takahasi:1974} as the integral 
can be rewritten as
\begin{equation}
 \int_0^\pi d\theta\sin^2\theta \rightarrow \int_{-1}^1 d\xi\sqrt{1-\xi^2},
\end{equation}
with $\xi\equiv\cos\theta$. This quadrature rule is well-suited for the occurring 
integrands, showing singular behavior at the boundaries of the integration region. 
Compared to a \textit{Gauss-Chebyshev quadrature} we need much less integration points 
to get the same - if not better - results for the angular integration.
With the following ansatz for the ghost propagator
\begin{equation}
 D_G(p) = -\frac{G(p^2)}{p^2}
\end{equation}
as well as for the gluon propagator
\begin{equation}
 D_{\mu\nu}(p) = \left(\delta_{\mu\nu} - \frac{p_\mu p_\nu}{p^2}\right)
 \frac{Z(p^2)}{p^2},
\end{equation}
the corresponding DSEs can be rewritten
\begin{eqnarray}
 G(x)^{-1} & = & \tilde Z_3 - \frac{g^2N_c}{(2\pi)^3}\int_0^{\Lambda^2}dy\;y\;
 G(y)\int_{-1}^1 d\xi\;(1-\xi)^{3/2}\,\frac{Z(z)}{z^2} \, ,\\
 Z(x)^{-1} & = & Z_3 + \frac{g^2N_c}{(2\pi)^3}\frac{1}{3x}\int_0^{\Lambda^2}dy\;
 \int_{-1}^1 d\xi\;(1-\xi^2)^{1/2}\,Q(x,y,z)\frac{\left(G(y)G(z)\right)^{-2-6\delta}}
 {\left(Z(y)Z(z)\right)^{3\delta}}\nonumber\\
           &   & \hspace{0.55cm} + \;\frac{g^2N_c}{(2\pi)^3}\frac{1}{3x}
	   \int_0^{\Lambda^2}dy\;\int_{-1}^1 d\xi\;(1-\xi^2)^{1/2}\,M(x,y,z)G(y)G(z),
\end{eqnarray}
where the integral kernels read\footnote{In the last line an additional term 
$15/4$ is added in order to cancel spurious quadratic divergencies introduced by the 
truncation. These divergencies would be absent in a full treatment. A pragmatic 
approach is to remove them by hand by adding an additional term. Furthermore, in the 
numerical implementation we optimized the kernels to avoid unnecessary division 
operations.}
\begin{eqnarray}
 M(x,y,z) & = & \frac{1}{z}
 \left(-\frac{1}{4}x+\frac{y}{2}-\frac{1}{4}\frac{y^2}{x}\right) 
 + \frac{1}{2} + \frac{1}{2}\frac{y}{x} - \frac{1}{4}\frac{z}{x} \, ,\\
 Q(x,y,z) & = & \frac{1}{z^2} 
 \left(\frac{1}{8}\frac{x^3}{y} + x^2 - \frac{9}{4}xy + y^2 + \frac{1}{8}\frac{y^3}{x}
 \right) \nonumber\\
          & + & \frac{1}{z} \left(\frac{x^2}{y} - 4(x+y) + \frac{y^2}{x}\right) - 
	  \left(\frac{9}{4}\frac{x}{y} + 4 + \frac{9}{4}\frac{y}{x}\right) \nonumber\\
  & + & z\left(\frac{1}{x} + \frac{1}{y}\right) + z^2\frac{1}{8xy} + \frac{15}{4}.
 \end{eqnarray}
To introduce a convenient notation the above equations are rewritten
\begin{eqnarray}
 G(x)^{-1} & = & \tilde Z_3 + \varPi_G(x) \, ,\\
 Z(x)^{-1} & = & Z_3 + \varPi_Z(x)\, ,
\end{eqnarray}
and a MOM scheme is applied in the next step
\begin{eqnarray}
 G(x)^{-1} & = & G(x_G)^{-1} + \varPi_G(x) - \varPi_G(x_G)\, \\
 Z(x)^{-1} & = & Z(x_Z)^{-1} + \varPi_Z(x) - \varPi_Z(x_Z).
\end{eqnarray}
In this renormalization scheme we subtract the equations at some squared momenta 
$x_G$ and $x_Z$, treating $G(x_G)^{-1}$ and $Z(x_Z)^{-1}$ as new parameters to fix the 
system. Furthermore, in the limits of very small/large external momenta $x$ the system 
can be solved analytically, where in the infrared region one finds\footnote{See 
Ref.~\cite{Fischer:2003zc} for a detailed analysis of the IR/UV behavior of the 
dressing functions.}
\begin{eqnarray}
 Z(x\ll 1) & = & A\,x^{2\kappa} \, ,\\
 G(x\ll 1) & = & B\,x^{-\kappa} \, ,
\end{eqnarray}
with $\kappa\approx0.595353$ and some general coefficients $A$ and $B$. 
These two coefficients are fixed by demanding that the analytical solution must 
coincide with the numerical solution at a specific matching point $\epsilon^2$ in the 
infrared regime. Similar to this, one can apply a logarithmic ansatz for the dressing 
functions in the ultraviolet region
 \begin{eqnarray}
  G(x\gg 1) & = & G_{UV} \left[\omega\ln\left(\frac{x}{x_{UV}}\right)+1\right]^\delta 
  \,\\
  Z(x\gg 1) & = & Z_{UV} \left[\omega\ln\left(\frac{x}{x_{UV}}\right)+1\right]^\gamma
  \, .
 \end{eqnarray}
The analytical treatment yields $\omega=11N_c\alpha(\mu^2)/12\pi$, where 
$\alpha(\mu^2) = g^2/4\pi$ is the coupling at the renormalization scale $\mu$. 
The constants $G_{UV}$ and $Z_{UV}$ are fixed by matching the numerical solutions at 
$x_{UV}=\Lambda^2$ and $\gamma=-13/22$ is the anomalous dimension of the gluon.

To allow to put in the following sections the focus on the numerical methods only  we
will present the numerical results for the dressing functions, the running coupling and
the Schwinger functions already here.

\subsection{Dressing Functions and Running Coupling}
\begin{figure}[ht!]
 \centering
 \includegraphics[scale=0.8]{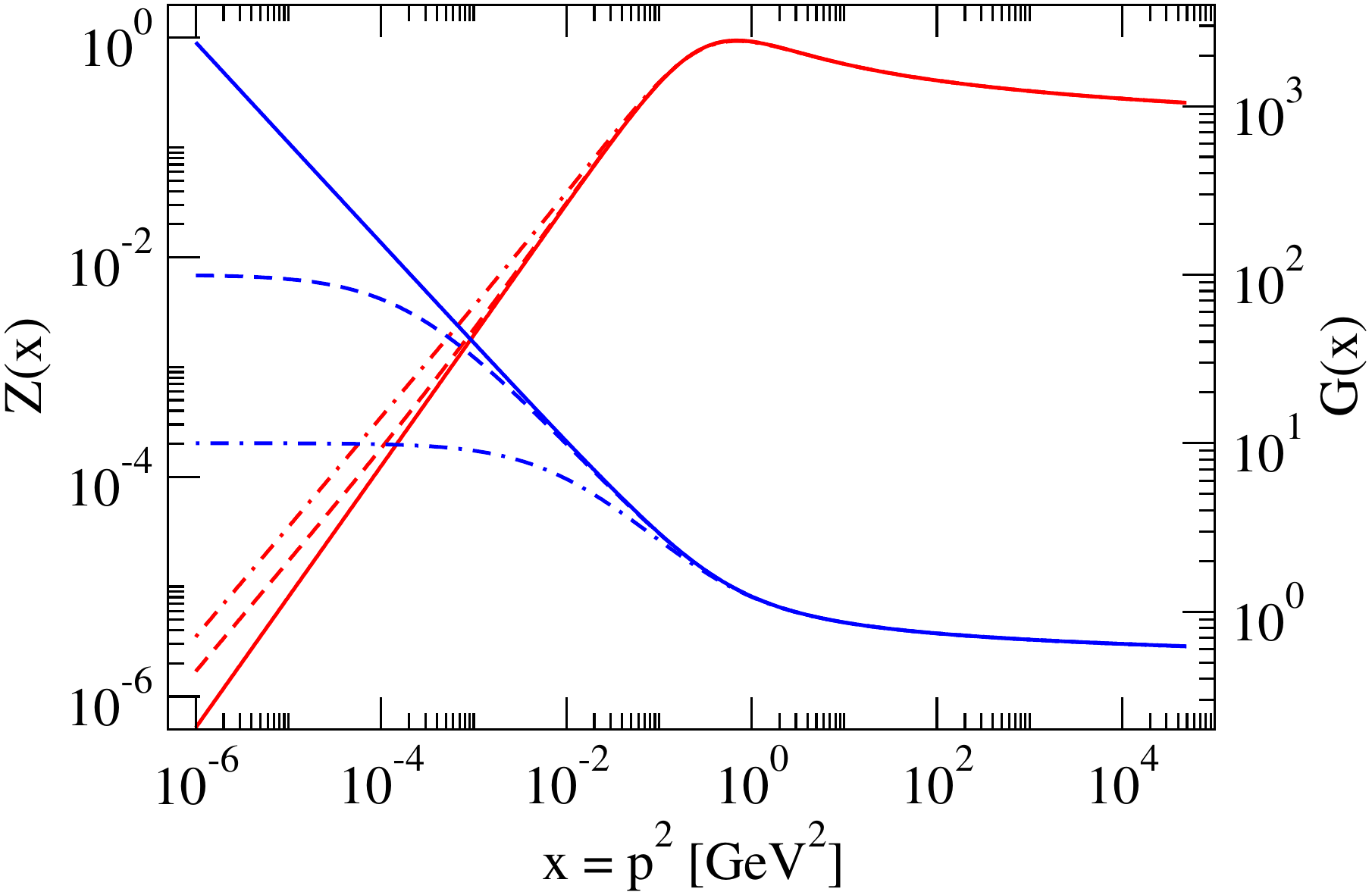}
 \caption{The scaling solution (solid line) as well as two decoupling solutions 
 (dashed lines) for the ghost and gluon propagator on a $log-log$ plot. For the 
 latter case, we used $G(x_G)^{-1}=0.01$ (dashed line) and $G(x_G)^{-1}=0.1$ 
 (dashed-dotted line).}
\label{fig:GhostGluonDressing}
\end{figure}

With an appropriate choice of $G(x_G)$ and $Z(x_Z)$ the system is fixed, where we use $x_G=0$ and 
$x_Z=\Lambda^2=5\times10^4\,GeV^2$ in our calculations. In 
Fig.~\ref{fig:GhostGluonDressing} the ghost and gluon dressing function is plotted, 
where $Z(x_Z)=0.256$ and $\alpha(\mu^2)=1$ is used\footnote{Note that these values 
are arbitrary in principle. Although, together with our choice for $\alpha(\mu^2)$ 
they yield the correct experimental value $\alpha(M_Z^2)=0.118$ for the running 
coupling, where $M_Z=91.2\,GeV$ is the Z-boson mass. The renormalization scale $\mu$ is 
implicitly fixed by specifying a value for $\alpha(\mu^2)$, where the renormalization 
condition $G^2(\mu^2)Z(\mu^2)=1$ is used.}.

\begin{figure}[ht!]
 \centering
 \includegraphics[scale=0.8]{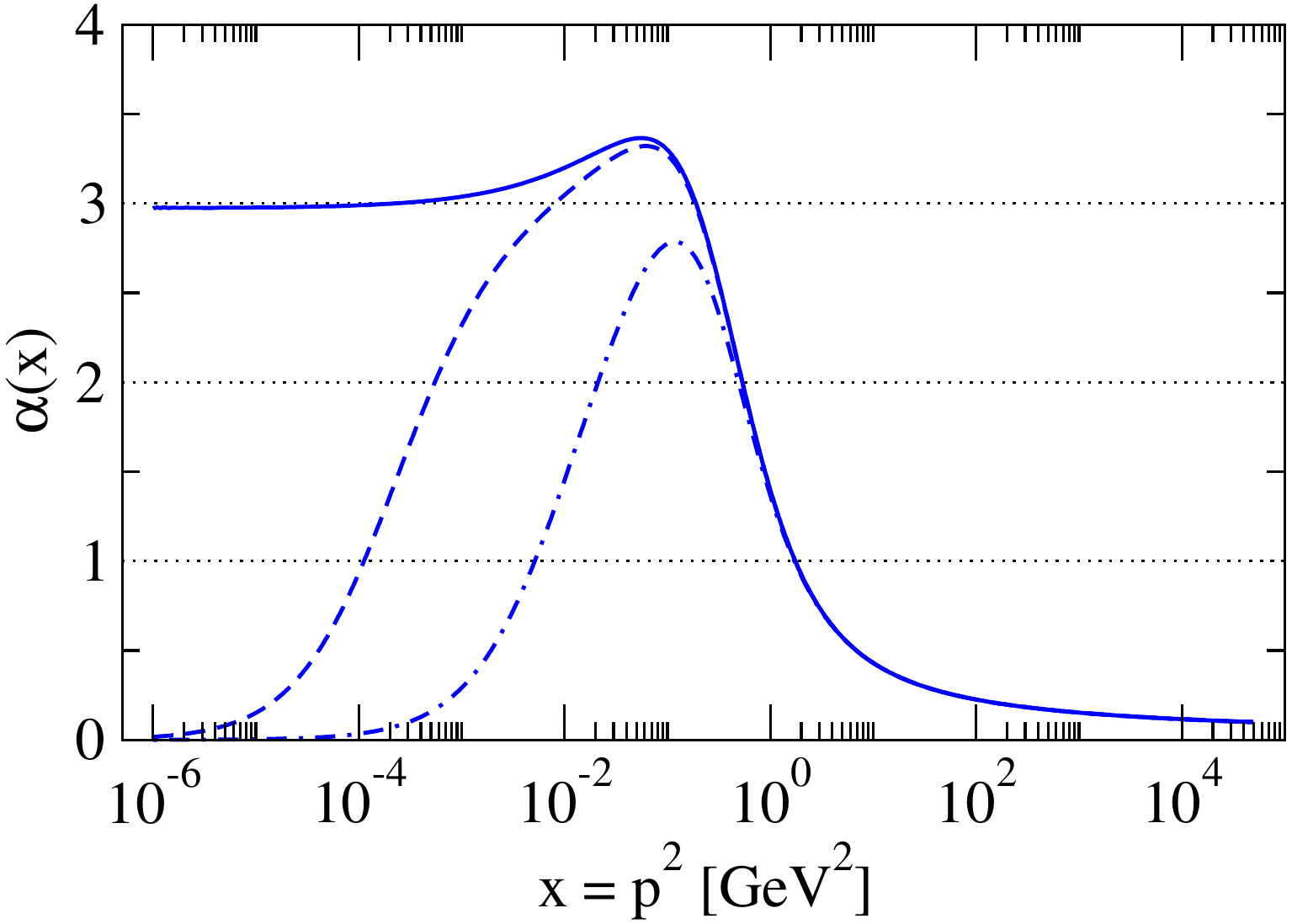}
 \caption{The non-perturbative running coupling $\alpha(x)$ on a $log-linear$ plot. 
 The scaling solution (solid line) is compared to the decoupling solution, where in the 
 latter case $G(x_G)^{-1}=0.01$ (dashed line) and $G(x_G)^{-1}=0.1$ 
 (dashed-dotted line) is used.}
\label{fig:Coupling}
\end{figure}

The choice of $G(x_G)^{-1}=0$ leads to an infrared singular ghost dressing function, 
whereas the corresponding gluon dressing function vanishes in this regime. This is the 
so-called \textit{scaling solution}. A non-vanishing value of $G(x_G)^{-1}>0$ yields 
a \textit{decoupling solution} which is indicated by the dashed/dashed-dotted lines. 
In Fig. \ref{fig:Coupling} the non-perturbative running coupling 
$\alpha(x) = \alpha(\mu^2)G^2(x)Z(x)$ is plotted.

The infrared fixed point is $\alpha(0)\approx8.915/N_C=2.972$ in the scaling case. 
In the decoupling case the running coupling vanishes in the infrared region which 
agrees with DSE calculations on a compact manifold \cite{Fischer:2005nf} and 
lattice calculations \cite{Sternbeck:2005tk}.

\subsection{Schwinger Functions}
\label{sec:SchwingerResults}
\begin{figure}[ht!]
 \centering
 \includegraphics[scale=0.7]{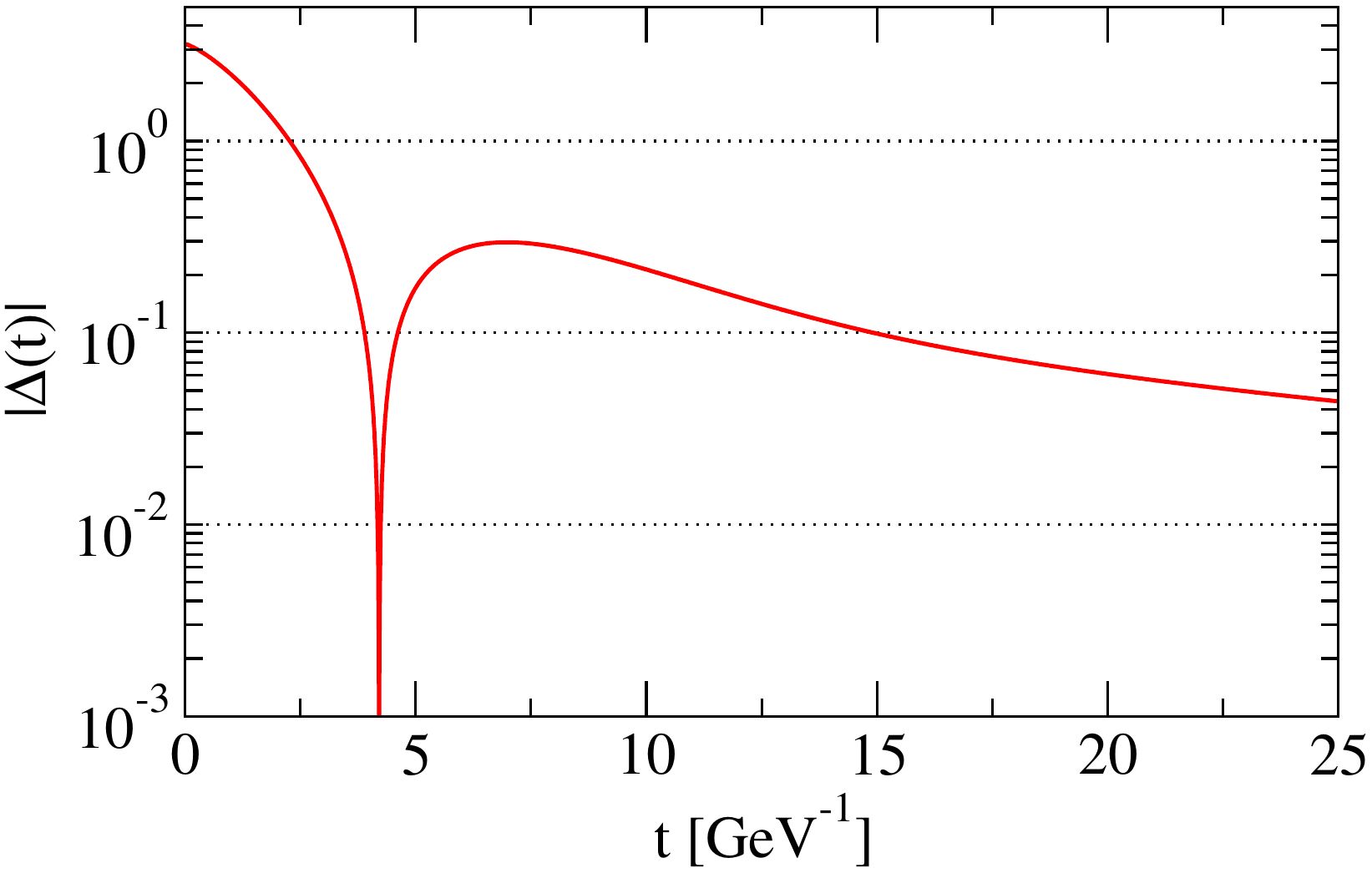}
 \caption{Displayed is the absolute value of the gluon propagator's Schwinger function 
 on a $log-linear$ plot. In accordance with \cite{Alkofer:2003jj} a zero slightly below 
 $t=5{\mathrm {GeV}}^{-1}$ is observed which indicates the violation of positivity at a 
 scale of roughly 1 fm.}
 \label{fig:SchwingerResult}
\end{figure}

Schwinger functions are a helpful tool when exploring the analytic structure of 
propagators. Although a detailed description is beyond the scope of this 
paper\footnote{A concise treatment can be, {\it e.g.}, found in  
refs.\ \cite{Haag:1992hx,Glimm:1987ng}.} we want to note the following:
Within the Euclidean formulation of quantum field theory, negative norm contributions 
to a  specific propagator correspond to a violation of the Osterwalder-Schrader 
axiom of \textit{reflection positivity} \cite{Osterwalder:1973dx}. Accordingly, 
this propagator does not possess a K\"allen-Lehman spectral representation and 
cannot describe a physical asymptotic state. Positivity violation on the level of 
propagators can be tested with the help of Schwinger functions defined as
\begin{equation}
\label{eq:Schwingerfunction}
 \Delta(t):=\frac{1}{\pi}\int_0^\infty d|p|\,\cos(t\,|p|)\sigma(p^2)\geq0,
\end{equation}
where $\sigma(p^2)$ is a scalar function extracted from the according 
propagator  which in case of the gluon is given by 
$\sigma(p^2)=Z(p^2)/p^2$ \cite{Alkofer:2003jj}.
In Fig.~\ref{fig:SchwingerResult} we show the Schwinger function for the gluon propagator
obtained from our numerical treatment of the ghost-gluon system.

\section{Outline of the Numerical Method}
\label{sec:numerics}
The renormalized system of coupled integral equations reads
\begin{eqnarray}
 G(x)^{-1} & = & G(x_G)^{-1} + \varPi_G(x) - \varPi_G(x_G),\\
 Z(x)^{-1} & = & Z(x_Z)^{-1} + \varPi_Z(x) - \varPi_Z(x_Z),
\end{eqnarray}
where $\varPi_G(x)$ and $\varPi_Z(x)$ denote the ghost and gluon self-energy terms, 
respectively. We now employ a Chebyshev expansion for the logarithms of the dressing 
functions
\begin{eqnarray}
 \ln G(x) & = & \frac{b_0}{2} + \sum_{j=1}^{N-1}b_jT_j(s(x)),\\
 \ln Z(x) & = & \frac{a_0}{2} + \sum_{j=1}^{N-1}a_jT_j(s(x)),
\end{eqnarray}
where $s$ is a suitable mapping function, which maps the $[-1,1]$ interval of the 
Chebyshev polynomials to the interval $[\epsilon^2,\Lambda^2]$. We use the one proposed 
in \cite{Atkinson:1997tu} which reads
\begin{equation}
 s(x) = \frac{\log_{10}(x/\Lambda\epsilon)}{\log_{10}(\Lambda/\epsilon)}.
\end{equation}
The values for the infrared matching point and the ultraviolet cutoff used in our 
calculations are given by $\epsilon^2=10^{-6}\,{\mathrm {GeV}}^2$ and 
$\Lambda^2 =5\times10^4\,{\mathrm {GeV}}^2$, respectively. 
Expanding the logarithm of the dressing functions leads to better results in the UV 
regime and to a significant reduction of Chebyshev polynomials needed in the 
expansion \cite{Maas:2005xh}. Furthermore, we split the radial integration according to
\begin{equation}
\label{eq:IntegrationInterval}
 \int_0^{\Lambda^2} \rightarrow \int_0^{\epsilon^2} + \int_{\epsilon^2}^x + 
 \int_x^{\Lambda^2}.
\end{equation}
In our numerical treatment the dressing functions $G(x)$ and $Z(x)$ are evaluated 
within the range $x\in[\epsilon^2,\Lambda^2]$. However, due to the appearance of the 
argument $z=x+y-2\sqrt{xy}\cos\theta\in[0,4\Lambda^2]$ extrapolated values are required. 
Thus, depending on the region, a function returns the appropriate values for $G(z)$ and 
$Z(z)$ by taking either the numerical or the respective analytical solution. 
We now get a non-linear system for the Chebyshev coefficients, such that the following 
equations have to be fulfilled for all external momenta $x$
\begin{equation}
\begin{split}
 g(x;\mathbf a, \mathbf b) & \equiv G(x)^{-1} - G(x_G)^{-1} - \varPi_G(x) 
 + \varPi_G(x_G) \;=\; 0 ,\\
 z(x;\mathbf a, \mathbf b) & \equiv Z(x)^{-1} - Z(x_Z)^{-1} - \varPi_Z(x) + 
 \varPi_Z(x_Z) \;=\; 0.
\end{split}
\end{equation}
Here, a vector notation for the Chebyshev coefficients is used
\begin{equation}
\begin{split}
 \mathbf a & \equiv (a_0,a_1,\ldots,a_{N-1})^T ,\\
 \mathbf b & \equiv (b_0,b_1,\ldots,b_{N-1})^T.
\end{split}
\end{equation}
In order to treat the unknown $2N$ Chebyshev coefficients, we evaluate the two 
equations at $N$ external momenta\footnote{In our calculations we use the mapped 
roots of the Chebyshev polynomials.} $x_i$
 \begin{eqnarray}
  g(x_i;\mathbf a,\mathbf b) & \equiv & G(x_i)^{-1} - G(x_G)^{-1} - \varPi_G(x_i) + 
  \varPi_G(x_G) \;=\; 0 ,\\
  z(x_i;\mathbf a,\mathbf b) & \equiv & Z(x_i)^{-1} - Z(x_Z)^{-1} - \varPi_Z(x_i) + 
  \varPi_Z(x_Z) \;=\; 0.
 \end{eqnarray}
According to \cite{Atkinson:1997tu}, a Newton-Raphson method is subsequently used to 
linearize the system.
It maps the non-linear system of equations for the 
Chebyshev coefficients to a linear system for the so-called Newton improvements 
which makes a matrix representation possible. During each iteration step new 
improvements are generated by a derivative of the functions $z$ and $g$ with 
respect to the $2N$ Chebyshev coefficients. These Newton improvements are then 
subtracted from the old coefficients in order to create a new set which is closer 
to the real solution, where for the initialization a starting guess is required. 
The two equations for the coefficients and their improvements reads
 \begin{eqnarray}
  a_j^{n+1} & = & a_j^n - \varrho\Delta a_j^{n+1} ,\\
  b_j^{n+1} & = & b_j^n - \varrho\Delta b_j^{n+1},
 \end{eqnarray}
where $n$ is the iteration step. Here, the last terms are additionally decorated with an
under-relaxation parameter $\varrho$. One advantage of the Newton method is that if the
system is close enough to the real solution, the method converges quadratically. This of
course depends on the starting guess. The under-relaxation parameter is used to relax the
system close to the real solution also from unlucky choices of the initial conditions.
After some iteration steps, this parameter can be set back to one in order to benefit
from the quadratic convergence rate of the method.
The Newton improvements $\Delta a$ and $\Delta b$ are described by a $2N\times2N$ set 
of linear equations 
 \begin{eqnarray}
  \dfrac{\partial z^n(x_i;\mathbf a,\mathbf b)}{\partial a_j}\Delta a_j^{n+1} + 
  \dfrac{\partial z^n(x_i;\mathbf a,\mathbf b)}{\partial b_j}\Delta b_j^{n+1} & = & 0, \\
  \dfrac{\partial g^n(x_i;\mathbf a,\mathbf b)}{\partial a_j}\Delta a_j^{n+1} + 
  \dfrac{\partial g^n(x_i;\mathbf a,\mathbf b)}{\partial b_j}\Delta b_j^{n+1} & = & 0.
 \end{eqnarray}
This linear system is represented by the following matrix acting on a vector for 
the $\Delta$'s
 \begin{equation} 
  \begin{pmatrix}
   \begin{matrix}
    \dfrac{\partial z(x_0)}{\partial a_0} & \hspace{-0.2cm}\ldots & 
    \hspace{-0.2cm}\dfrac{\partial z(x_0)}{\partial a_{N-1}} \\
    \vdots & & \vdots \\
    \dfrac{\partial z(x_{N-1})}{\partial a_0} & \hspace{-0.2cm}\ldots & 
    \hspace{-0.2cm}\dfrac{\partial z(x_{N-1})}{\partial a_{N-1}} \\
   \end{matrix} &
   \begin{matrix}
    \dfrac{\partial z(x_0)}{\partial b_0} & \hspace{-0.2cm}\ldots & 
    \hspace{-0.2cm}\dfrac{\partial z(x_0)}{\partial b_{N-1}} \\
    \vdots & & \vdots \\
    \dfrac{\partial z(x_{N-1})}{\partial b_0} & \hspace{-0.2cm}\ldots & 
    \hspace{-0.2cm}\dfrac{\partial z(x_{N-1})}{\partial b_{N-1}} \\
   \end{matrix} \\\\
   \begin{matrix}
    \dfrac{\partial g(x_0)}{\partial a_0} & \hspace{-0.2cm}\ldots & 
    \hspace{-0.2cm}\dfrac{\partial g(x_0)}{\partial a_{N-1}} \\
    \vdots & & \vdots \\
    \dfrac{\partial g(x_{N-1})}{\partial a_0} & \hspace{-0.2cm}\ldots & 
    \hspace{-0.2cm}\dfrac{\partial g(x_{N-1})}{\partial a_{N-1}} \\
   \end{matrix} &
   \begin{matrix}
    \dfrac{\partial g(x_0)}{\partial b_0} & \hspace{-0.2cm}\ldots & 
    \hspace{-0.2cm}\dfrac{\partial g(x_0)}{\partial b_{N-1}} \\
    \vdots & & \vdots \\
    \dfrac{\partial g(x_{N-1})}{\partial b_0} & \hspace{-0.2cm}\ldots & 
    \hspace{-0.2cm}\dfrac{\partial g(x_{N-1})}{\partial b_{N-1}} \\
   \end{matrix} \\
   \end{pmatrix}
   \begin{pmatrix}
    \Delta a_0 \\
    \vdots \\
    \Delta a_{N-1} \\
    \Delta b_0 \\
    \vdots \\
    \Delta b_{N-1} \\
   \end{pmatrix} = 
   \begin{pmatrix}
    z(x_0) \\
    \vdots \\
    z(x_{N-1}) \\
    g(x_0) \\
    \vdots \\
    g(x_{N-1}) \\
   \end{pmatrix} \stackrel{!}{=}
   \begin{pmatrix}
    0 \\
    \vdots \\
    0 \\
    0 \\
    \vdots \\
    0 \\
  \end{pmatrix}.
 \end{equation}
After the system iterated several times and convergence is achieved, the solutions 
have to vanish. Thus one needs to generate and invert the Newton matrix in order 
to get a new set of Newton improvements during each iteration step. A detailed 
description on the implementation of the numerical method outlined above will be 
given after some general remarks on GPU programming.
\section{GPU Calculations using CUDA\texttrademark}
\label{sec:GPUsection}
During the last years, programmable Graphic Processor Units (GPUs) became more and 
more important in scientific and engineering high performance computing. Due to their 
multi-threaded manycore processor architecture, they are perfectly well-suited in 
dealing with compute-intensive, parallel programs. There are several programming models 
available like OpenCL\texttrademark or DirectCompute\texttrademark for instance. We use 
CUDA\texttrademark for the numerical implementation of our problem, the parallel 
computing model provided by NVIDIA$^{\textregistered}$.  

\subsection{Kernels, Blocks and Threads} 

A CUDA\texttrademark code runs on the host CPU as a serial program in which the compute
intensive parts are condensed into one our more kernel functions to be performed in
parallel on the GPU device.
In order to be scalable, the
kernels are structured into a three-level hierarchy of threads. The top level is
represented by a grid of
thread-blocks with up to 1024 threads acting within
one block. Thread-blocks represent the second hierarchy level and contain the same number
of threads in each case. A single thread stands at the lowest hierarchy level and
represents the basic building block of a kernel function. The distribution of thread-blocks 
to the different device multiprocessors is managed by the hardware and there is no possibility 
to decide which block is performed on which multiprocessor at a given time. For that reason, 
blocks have to be independent from each other.

A single multiprocessor which operates on a specific block executes several threads in
parallel, grouped to a so-called warp. The individual threads in a warp have the same
start address in the program but can act on/with different data and are in principle free
to branch and follow different execution paths. In this case each branch path is executed
sequentially (warp divergence), where at the end the threads converge back to the
original execution path and the multiprocessor can handle the next warp. Warp divergence
has a considerable effect on the run time of the program and best performance is achieved
if all threads of a warp agree on their particular execution path. 

Each multiprocessor owns a specific number of registers. These memory spaces are
dynamically assigned to the corresponding threads currently running on the
multiprocessor. Whereas each thread can operate on its specific register space, there
is also the possibility that threads within a block can interchange data via a very fast
shared memory. Communication between different blocks in a grid can only occur via
a global memory space which is slower than the shared memory. The global memory can be
accessed by all the threads within the kernels as well as by the host to read/write the
corresponding data. Although global memory access is still roughly ten times faster than
the main memory access of a conventional CPU, it is favorable to minimize the
communication of threads with the global memory during a kernel call.
If the memory access is not performed in a coalesce way
\cite{NVIDIA:programming_guide,NVIDIA:design_guide} it can be that some threads already
operate on their specific data while others have to wait for data arriving from the
global memory.

In general the optimization of a CUDA\texttrademark program benefits from several tasks.
First the communication between the host and the device as well as out-of-order accesses
to the global memory from kernels have to be minimized. Furthermore, the overall
performance of the program also depends on the number of blocks and the blocksize.
Running only one block per multiprocessor can
force the specific processor to idle because of
latencies in memory access and/or block synchronization. Launching several blocks
decreases register and shared memory resources available to a single thread-block. The
blocksize strongly depends on the workload each thread has to perform and on the number
of registers available per block. It can very well happen that the possible number of
threads is below the maximal number of threads given in the device specifications simply
because of the heavy register usage of compute intensive threads.

\subsection{Implementation on a Graphics Device with CUDA\texttrademark}
The central task of the code is to generate the Jacobian matrix for the Newton
improvements. Therefore we split the matrix into sub-matrices which are distributed onto
four different kernel functions 
 \begin{equation}
  \mathcal J = 
  \begin{pmatrix}
   \textcolor{red}{\begin{matrix}
   \dfrac{\partial z(x_0)}{\partial a_0} & \hspace{-0.3cm}\ldots & 
   \hspace{-0.3cm}\dfrac{\partial z(x_0)}{\partial a_{N-1}} \\
   \vdots & \hspace{-0.3cm}Kernel\;1 & \hspace{-0.3cm}\vdots \\
   \dfrac{\partial z(x_{N-1})}{\partial a_0} & \hspace{-0.3cm}\ldots & 
   \hspace{-0.3cm}\dfrac{\partial z(x_{N-1})}{\partial a_{N-1}}
  \end{matrix}} & \hspace{-0.2cm}
   \textcolor{blue}{\begin{matrix}
   \dfrac{\partial z(x_0)}{\partial b_0} & \hspace{-0.3cm}\ldots & 
   \hspace{-0.3cm}\dfrac{\partial z(x_0)}{\partial b_{N-1}} \\
   \vdots & \hspace{-0.3cm}Kernel\;2 & \hspace{-0.3cm}\vdots \\
   \dfrac{\partial z(x_{N-1})}{\partial b_0} & \hspace{-0.3cm}\ldots & 
   \hspace{-0.3cm}\dfrac{\partial z(x_{N-1})}{\partial b_{N-1}}
  \end{matrix}}\\\\
   \textcolor{magenta}{\begin{matrix}
   \dfrac{\partial g(x_0)}{\partial a_0} & \hspace{-0.3cm}\ldots & 
   \hspace{-0.3cm}\dfrac{\partial g(x_0)}{\partial a_{N-1}} \\
   \vdots & \hspace{-0.3cm}Kernel\;3 & \hspace{-0.3cm}\vdots \\
   \dfrac{\partial g(x_{N-1})}{\partial a_0} & \hspace{-0.3cm}\ldots & 
   \hspace{-0.3cm}\dfrac{\partial g(x_{N-1})}{\partial a_{N-1}}
  \end{matrix}} & \hspace{-0.2cm}
   \textcolor{cyan}{\begin{matrix}
   \dfrac{\partial g(x_0)}{\partial b_0} & \hspace{-0.3cm}\ldots & 
   \hspace{-0.3cm}\dfrac{\partial g(x_0)}{\partial b_{N-1}} \\
   \vdots & \hspace{-0.3cm}Kernel\;4 & \hspace{-0.3cm}\vdots \\
   \dfrac{\partial g(x_{N-1})}{\partial b_0} & \hspace{-0.3cm}\ldots & 
   \hspace{-0.3cm}\dfrac{\partial g(x_{N-1})}{\partial b_{N-1}}
  \end{matrix}}
  \end{pmatrix}.
 \end{equation}
These kernels can now be launched on the available GPU devices. Once the matrix 
elements are derived, a data transfer to the host takes place where the inversion 
of the matrix is performed. As the size of the matrix is rather small, the inversion 
can be done with standard LU decomposition routines on the host. The time usage of this 
procedure is negligible compared to the generation of the matrix. On the particular 
device the sub-matrix is split into blocks of size $N$
\begin{equation}
 \begin{split}
  &\textcolor{red}{\left(
  \begin{matrix}
   \textcolor{red}{\dfrac{\partial z(x_0)}{\partial a_0}} & 
   \textcolor{blue}{\dfrac{\partial z(x_0)}{\partial a_1}} & 
   \textcolor{red}{\ldots} & 
   \textcolor{magenta}{\dfrac{\partial z(x_0)}{\partial a_{N-1}}} \\
   \textcolor{red}{\dfrac{\partial z(x_1)}{\partial a_0}} & 
   \textcolor{blue}{\dfrac{\partial z(x_1)}{\partial a_1}} & 
   \textcolor{red}{\ldots} & 
   \textcolor{magenta}{\dfrac{\partial z(x_1)}{\partial a_{N-1}}} \\
   \vdots & \textcolor{blue}{\vdots} & GPU 1 & 
   \textcolor{magenta}{\vdots} \\
   \textcolor{red}{\dfrac{\partial z(x_{N-1})}{\partial a_0}} & 
   \textcolor{blue}{\dfrac{\partial z(x_{N-1})}{\partial a_1}} & 
   \textcolor{red}{\ldots} & 
   \textcolor{magenta}{\dfrac{\partial z(x_{N-1})}{\partial a_{N-1}}}
  \end{matrix}\right)}. \\
  &\hspace{0.55cm}\begin{matrix}
   \textcolor{red}{\uparrow} & \hspace{0.35cm}\textcolor{blue}{\uparrow} & 
   \hspace{0.55cm} & \hspace{0.45cm}\textcolor{magenta}{\uparrow} \\ 
   \textcolor{red}{Block\;1} & \hspace{0.35cm}\textcolor{blue}{Block\;2} & 
   \hspace{0.55cm}\textcolor{red}{\ldots} & \hspace{0.45cm}\textcolor{magenta}{Block\;N}
  \end{matrix}
 \end{split}
\end{equation}
These blocks are completely independent, {\it i.e.}, no data transfer or communication 
with the host or between the threads in the block is needed. In addition, 
the derivatives can be performed by two threads
\begin{equation}
\label{eq:derivatives}
 \dfrac{\partial z(x_i)}{\partial a_j} = 
 \dfrac{z(x_i;\mathbf a|_{a_j+\epsilon_j},\mathbf b) - z(x_i;\mathbf a|_{a_j-\epsilon_j},
 \mathbf b)}{2\epsilon_j},
\end{equation}
where finally one of the two threads has to sum up and divide the result by 
2$\epsilon_j$. In our numerical simulation we use $\epsilon_j \approx 10^{-5} a_j$ 
as the derivatives are symmetric. Note that in each block the same derivatives are 
used such that the corresponding vectors can be pre-calculated in parallel during 
each iteration step and stored as an array which is, in case of the matrix part treated 
in Eq. \eqref{eq:derivatives}, of the form
\begin{equation}
 \left(\begin{matrix}
  \textcolor{red}{a_0 + \epsilon_0} & \textcolor{red}{a_0}              
  & \textcolor{red}{\ldots} & \textcolor{red}{a_0}    \\ 
  \textcolor{red}{a_1}              & \textcolor{red}{a_1 + \epsilon_1} 
  & \textcolor{red}{\ldots} & \textcolor{red}{a_1}    \\ 
  \textcolor{red}{\vdots}           & \textcolor{red}{\vdots}           
  & \textcolor{red}{\ddots} & \textcolor{red}{\vdots} \\ 
  \textcolor{red}{a_{N-1}}          & \textcolor{red}{a_{N-1}}         
   & \textcolor{red}{\ldots} & \textcolor{red}{a_{N-1} + \epsilon_{N-1}} \\ 
  \hline
  \textcolor{blue}{a_0 - \epsilon_0} & \textcolor{blue}{a_0}              
  & \textcolor{blue}{\ldots} & \textcolor{blue}{a_0}    \\ 
  \textcolor{blue}{a_1}              & \textcolor{blue}{a_1 - \epsilon_1} 
  & \textcolor{blue}{\ldots} & \textcolor{blue}{a_1}    \\ 
  \textcolor{blue}{\vdots}           & \textcolor{blue}{\vdots}           
  & \textcolor{blue}{\ddots} & \textcolor{blue}{\vdots} \\ 
  \textcolor{blue}{a_{N-1}}          & \textcolor{blue}{a_{N-1}}          
  & \textcolor{blue}{\ldots} & \textcolor{blue}{a_{N-1} - \epsilon_{N-1}}
 \end{matrix}\right).
\end{equation}
These arrays are generated on the GPU devices according to their assigned sub-matrix. 
Subsequently, the blocks can load their specific derivative vector into the shared 
memory with respect to their block and thread IDs. Our code is separated into an 
initialization part running on the host and a main part which is performing the 
generation of the Newton matrix. The first part deals with the memory management, 
the initialization of the Chebyshev coefficients as well as the transfer of the 
corresponding data to the different devices. Note that the weights and nodes for the 
quadratures, which are generated on the host, are stored within the constant memory of 
the devices. The main part is split up into four kernel functions for the four different
 domains of the Newton matrix as well as one kernel function for the generation of the 
 solutions vector. These kernels are launched in parallel on the specific devices using 
 the \textit{cudaSetDevice} instruction. In our numerical implementation using OpenMPI each 
process invokes a GPU device according to its specific process ID
\footnote{We note that the usage of a separate GPU device for the 
  generation of the solutions vector is less efficient. As this step requires an 
  additional compute node, the benefits of the additional resources are most probably 
  spoiled by the expensive communication paths between the two machines as there are 
  only four GPU devices per machine. Here, the overall performance of the code is 
  improved if one of the four MPI processes, in our case the master process, performs 
  the calculation of the solutions vector as well as one ghost part of the matrix. 
  The master process collects the data from the other MPI processes via a 
  \textit{MPI\_Gather} operation such that in this setting no data transfer is needed. 
  We note that using this configuration also three GPUs would be enough, as the run 
  time of the two processes performing the ghost parts is comparable to the run time 
  of a single process performing one gluon part of the matrix. Nevertheless, the 
  following performance tests were carried out on four GPUs.}. 
  After the kernels generated their particular part of the matrix a data transfer to 
  the host is performed where the LU decomposition/substitution takes place. The 
  generation of the derivatives is performed in parallel by an additional kernel 
  in the beginning of each iteration step. In the following we show performance 
  results of the serial code and compare with a parallelized version using OpenMPI 
  as well as CUDA\texttrademark.
  
\section{Performance Results}
\label{sec:performance}

We tested various versions of the numerical method outlined in Sec. \ref{sec:numerics}.
The easiest approach is to perform the calculations on a single CPU device\footnote{The 
numerical problem discussed here can be solved quite fast on a single CPU, see table 1. The
current (described) calculations provided in table 1 serve, however, as an important test case 
because in the near
future much more involved truncation schemes of DSEs will be investigated.}. Subsequently,
we parallelized the code using OpenMPI which is, due to the nature of the problem, a
straightforward task. Each element of a Newton sub-matrix is generated in blocks
according to the number of processes launched, starting from the top left entry. Here,
only the master process has to allocate memory for the whole matrix since the other
processes are calculating just the elements. Since the communication load is negligible,
this procedure scales quite well with the number of processors available. The final step
is the GPU implementation, where we additionally compare a single and a multi GPU
version\footnote{For an accurate performance test we use rather large workloads and we
furthermore employ the fit functions proposed in \cite{Fischer:2003zc} to initiate the
system setting the under-relaxation parameter $\varrho=1$ in this case. Using this setup
the system iterates four times, where we implemented a cross sum check over all Chebyshev
coefficients with an epsilon of $\epsilon=10^{-6}$.}.

\vspace*{0.2cm}
For the numerical calculations the following hardware is used:
\begin{itemize}
 \vspace*{-0.2cm}
 \item Intel Core2 Quad Processor Q9400 @ 2.66GHz (no Hyper-Threading capabilities),
 \vspace*{-0.2cm}
 \item Intel Xeon Six-Core X5650 @ 2.67GHz,
 \vspace*{-0.2cm}
 \item NVIDIA Geforce GTX 560 Ti 448 Cores,
 \vspace*{-0.2cm}
 \item NVIDIA Tesla C2070.
 \vspace*{-0.1cm}
\end{itemize}
The Xeons are additionally connected via 40Gb/s QDR InfiniBand, whereas for the Core2s a
standard Gigabit-Ethernet LAN is used. The following table  shows performance results
obtained on the specific hardware using the different numerical implementations, where
the value in brackets denotes the number of cores/devices used for the corresponding
calculation. By varying the number of radial integration nodes and Chebyshev polynomials, six
different work-loads are employed.

\begin{table}[ht!]
 \centering
 \begin{tabular}{ | c || c | c | c | c | c | c | c | c | }
  \hline
  Nodes & Q(1) & Q(12) & Xe(1) & Xe(12) & GTX(1) & Tesla(1) & Tesla(4) & \textbf{Speed-up} \\
  \hline
  \hline
  32 & 1.63 & 0.14 & 3.75 & 0.42 & 0.08 & 0.07 & 0.04 & \textbf{94/11} \\
  64 & 3.27 & 0.29 & 7.34 & 0.86 & 0.13 & 0.11 & 0.06 & \textbf{122/14} \\
  \hline
  \hline
  64 & 14.80 & 1.30 & 23.45 & 2.05 & 0.27 & 0.23 & 0.10 & \textbf{235/21} \\
 128 & 29.58 & 2.58 & 48.38 & 4.25 & 0.53 & 0.45 & 0.15 & \textbf{323/28} \\
 \hline
 \hline
  64 &  54.32 & 4.70 &  74.22 &  6.40 & 0.58 & 0.40 & 0.15 & \textbf{495/43} \\
 128 & 119.33 & 9.39 & 152.05 & 13.08 & 1.15 & 0.75 & 0.27 & \textbf{563/48} \\
  \hline
 \end{tabular}
 \caption{The run time of the code in minutes using 36/60/96 Chebyshev polynomials 
 (upper/middle/lower part of the table) for each dressing function. For the radial 
 integral 32/64/128 Gauss-Legendre nodes are employed within each of the three integration 
 intervals, cf. eq.~\eqref{eq:IntegrationInterval}.}
 \label{tab:performance}
\end{table}

\begin{figure}[ht!]
 \centering
 \includegraphics[scale=0.69]{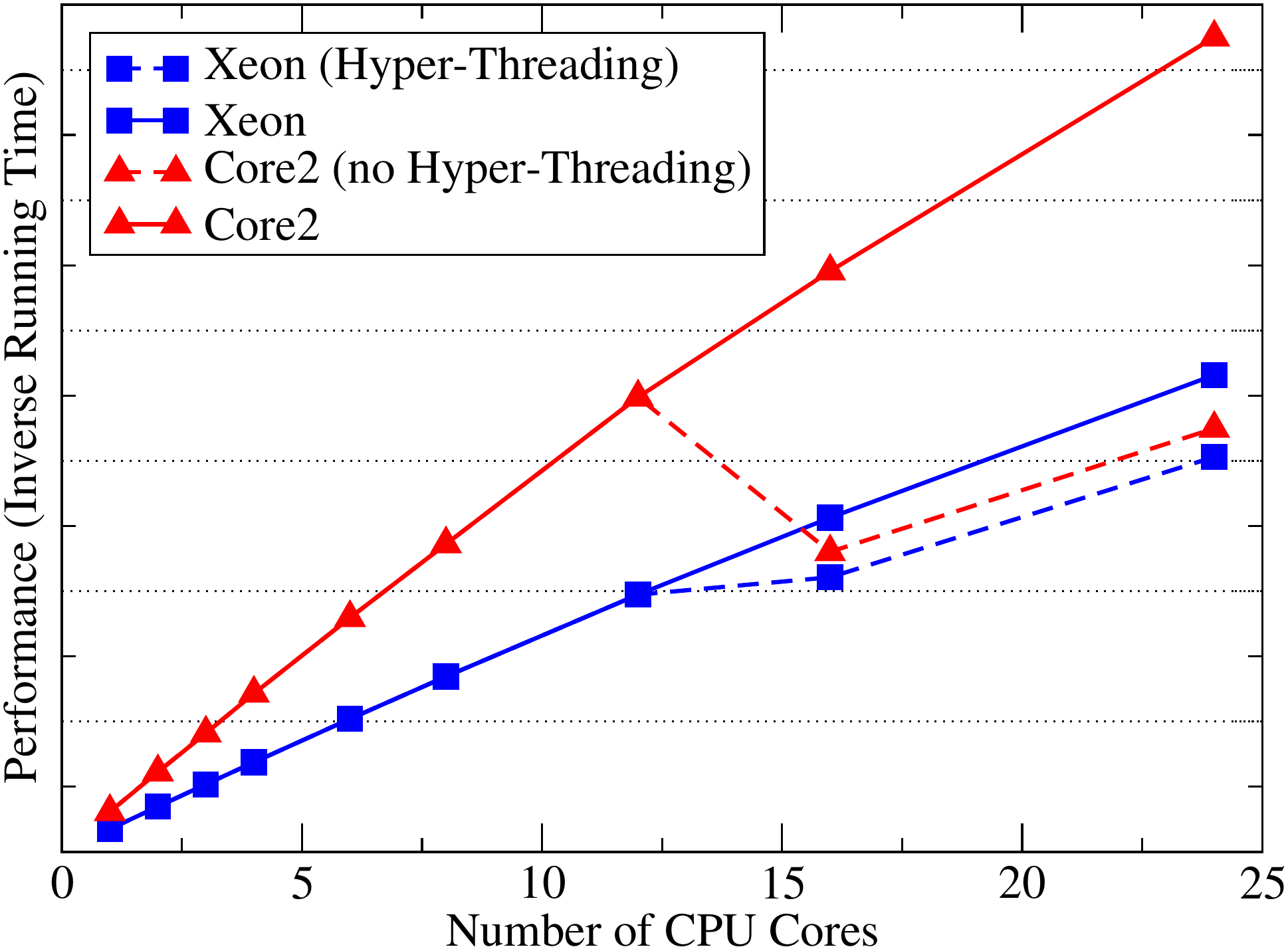}
 \caption{The expected linear performance gain of the MPI code when the number of 
 CPU cores is increased (solid lines). Whereas in a direct comparison of the two 
 architectures the Xeons are slower, Intel's Hyper-Threading technology increases the
 efficiency of the server CPUs by approximately 40\% (blue dashed line). Here, 16/24 threads are
 launched on twelve CPU cores. The Core2s show a considerable performance 
 break down in this case since Hyper-Threading is not available on these devices (red dashed line).}
 \label{fig:MPIscaling}
\end{figure}

The upper/middle/lower part of Tab. \ref{tab:performance} represents the performance test 
using 36/60/96 Chebyshev polynomials for each dressing function, where the entries denote 
the run time of the code in minutes. Here, the approximate speed-ups are measured between 
the slowest and the fastest implementation on the one hand (first number) and within 
a more equitable setup by comparing twelve Xeons with four Teslas (second number). 
With increasing work-load, the GPUs perform successively better and we would like to emphasize 
that already with the relatively low-priced consumer card impressive speed-ups can be obtained. 
In Fig. \ref{fig:MPIscaling} we show the scaling of the MPI code when the number of CPU cores is
increased (for both the Core2s and the Xeons). Here 128 radial integration points are used 
within each integration region as well as 48 Chebyshev polynomials for each dressing function. 

\begin{figure}[ht!]
 \centering
 \includegraphics[scale=0.69]{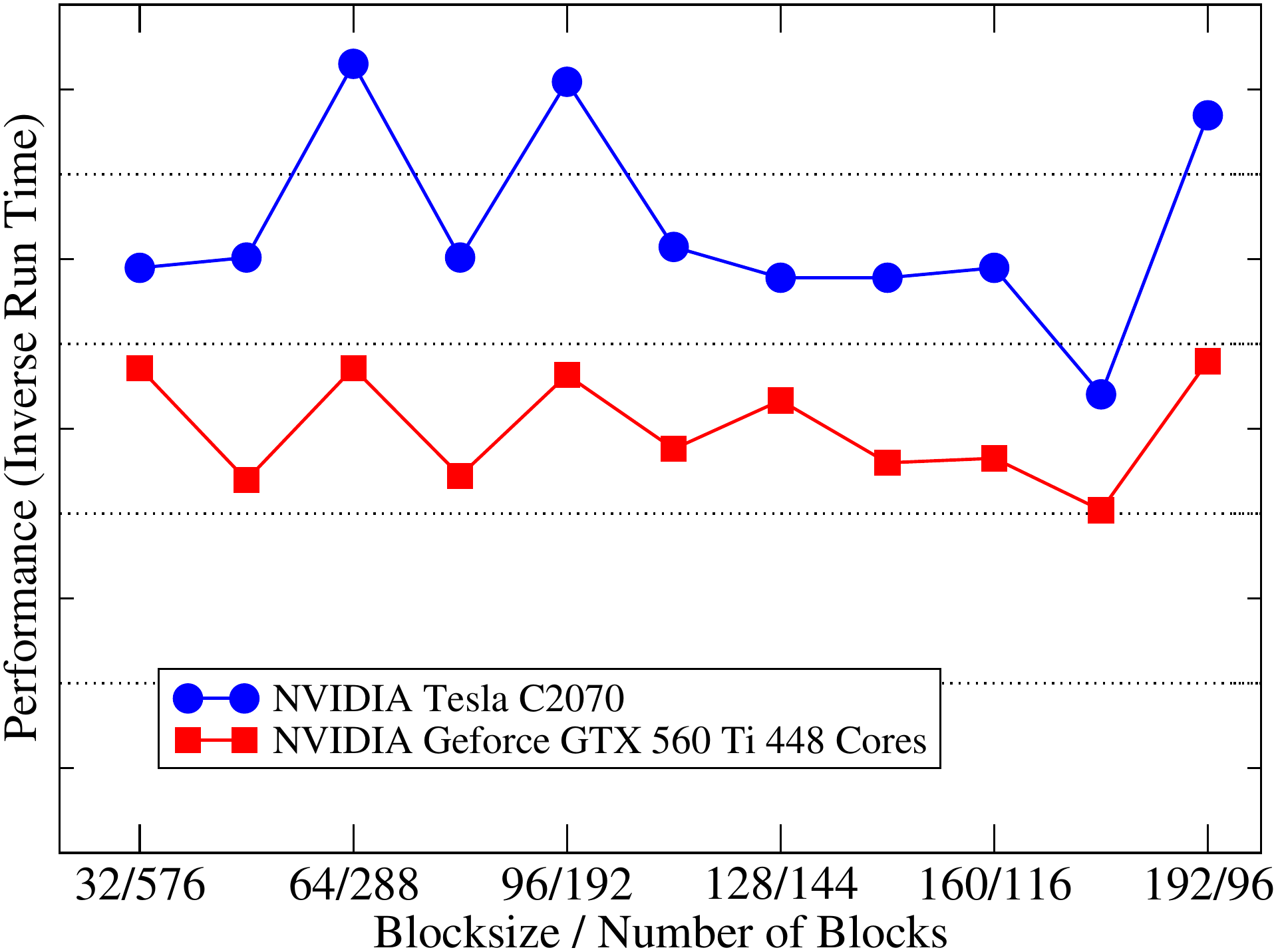}
 \caption{The impact of the blocksize on the performance for test runs on single GPUs
 using 96 Chebyshev polynomials and 128 radial integration points.
 The axes label denotes the blocksize and the number of blocks, respectively
 (note that two threads are performing a single matrix entry as mentioned 
 in Sec. \ref{sec:GPUsection}).
 One can see some slight deviations in the overall behavior of the different types of GPUs but in
 any case choosing the blocksize to be a multiple of a warp leads to optimal performance. 
 Furthermore, these effects become noticeable only for rather large workloads. Thus, a
 blocksize equal to the number of Chebyshev polynomials (and equal to a multiple of a warp) 
 is a convenient choice for almost all practical calculations.}
 \label{fig:BlockSizeDependence}
\end{figure}

Both measurements show the expected linear scaling behavior (solid lines). Launching more processes 
than CPU cores available (dashed lines) results in a serious performance drop of the Core2s. Due to 
Intel's Hyper-Threading technology, a single Xeon core can operate on two threads 
concurrently which results in a performance loss of only  20\% compared to a test run 
using the corresponding number of real cores. In Fig. \ref{fig:BlockSizeDependence} the blocksize
dependence of the CUDA\texttrademark code for the two different types of GPUs is shown.

Let us finally comment on the calculation of the Schwinger function. The integral 
in eq.~\eqref{eq:Schwingerfunction} is easily implemented on a GPU device since
it is a generalized scalar product and can be treated using standard techniques
\cite{NVIDIA:programming_guide}. The kernels are decomposed into a one-dimensional grid
of thread-blocks for the Euclidean time steps $\delta t$, where we used strides
\cite{NVIDIA:design_guide} to reduce the results in the end. However, the main
improvements can be obtained from a parallelized generation of the nodes and weights. For
the computation a (mapped) Gauss-Legendre quadrature rule is used.
The following table shows performance results where the entries denote the run time in
seconds\footnote{Here, the speed-ups are measured between the GPUs and the Xeons and we
used $2^{17}$ Gauss-Legendre nodes. Also in this case an OpenMPI version is possible in
principle. However, the usage of multiple GPUs is not reasonable due to the reduced
complexity of this problem.}.

\begin{table}[ht!]
 \centering
 \begin{tabular}{ | l || c | c | c | c | }
  \hline
                      & Q(1) & Xe(1) & GTX(1) & Tesla(1) \\
  \hline
  \hline
  sequential          & 213.5 & 226.9 &     &     \\
  \hline
  CUDA\texttrademark  &       &       & 4.1 & 3.9 \\
  \hline
  \hline
  \textbf{speed-up}   &  &  & \textbf{55} & \textbf{58} \\
  \hline
 \end{tabular}
 \caption{Performance results for an evaluation of the integral in
 eq.~\eqref{eq:Schwingerfunction}.}
 \label{tab:performance2}
\end{table}

\section{Conclusions}
\label{sec:conclusions}

We performed a numerical analysis of the ghost-gluon Dyson-Schwinger equations (DSEs) of
Yang-Mills theory. The truncated system of non-linear integral equations was solved with
the help of a Chebyshev expansion for the dressing functions using subsequently a
Newton-Raphson method to obtain a linear system. Here, the methods are ideally suited for
an SIMD architecture as the problem decomposes into independent parts. The
parallelization of the system was performed using OpenMPI and CUDA\texttrademark. 
By comparing the two parallelization strategies we demonstrated the computational advantage 
of GPUs for this problem. Compared to a sequential version we obtained speed-ups of approximately 
two orders of magnitude already with a single consumer GPU. The presented results
demonstrate convincingly the benefits of modern GPU devices in DSE calculations, and the
proposed solution strategy offers a helpful toolbox. 

Last but not least, the generalization to larger systems is straightforward since
additional DSEs can be incorporated by extending the Newton matrix with the corresponding
derivatives. In this respect we provided a basis for on-going and future computations 
which uses the Yang-Mills DSE system as input. Here, the GPU version does and is expected
to perform successively better with increasing workload.

\section*{Acknowledgements}

We thank Markus Q. Huber and Manfred Liebmann for valuable discussions. This work was
supported by the \textit{Research Core Area ``Modeling and Simulation''} of the
Karl-Franzens University Graz and by the Austrian Science Fund (FWF DK W1203-N16).





\goodbreak



\end{document}